\documentclass[useAMS,usenatbib]{mn2e}

\usepackage{graphicx}
\usepackage{amsmath}
\usepackage{amssymb}
\usepackage{natbib}

\newcommand{\apj}{Ap.J.}
\newcommand{\apjs}{Ap.J.S.}
\newcommand{\aj}{A.J.}
\newcommand{\mnras}{MNRAS}
\newcommand{\aap}{A\&A}

\def\trh0{t_{rh}(0)}

\newcommand{\be}{\begin{equation}}
\newcommand{\ee}{\end{equation}}

\def\apgt{\ {\raise-.5ex\hbox{$\buildrel>\over\sim$}}\ }
\def\aplt{\ {\raise-.5ex\hbox{$\buildrel<\over\sim$}}\ }

\title[Triple Stars with and without a Pulsar in Star Clusters]
{Predictions for Triple Stars with and without a Pulsar in Star
Clusters} 

\author[Trenti et al.]{Michele Trenti$^{1}$\thanks{E-mail
addresses: trenti@stsci.edu (MT); sransom@nrao.edu (SR); piet@ias.edu
(PH); d.c.heggie@ed.ac.uk (DCH)}, Scott Ransom$^{2}$\footnotemark[1],
Piet Hut$^{3}$\footnotemark[1] and Douglas
C. Heggie$^{4}$\footnotemark[1]\\ $^{1}$Space Telescope Science
Institute, Baltimore, MD 21210, U.S.A. \\ $^{2}$National Radio Astronomy
Observatory, 520 Edgemont Rd., Charlottesville, VA, 22903, U.S.A.\\
$^{3}$Institute for Advanced Study, Princeton, NJ 08540, U.S.A. \\
$^{4}$School of Mathematics and Maxwell Insitute of Mathematical
Sciences, University of Edinburgh, King's
Buildings, \\ Edinburgh EH9 3JZ, Scotland, U.K.}

\begin{document}

\date{Accepted ; Received ; in original form }

\pagerange{\pageref{firstpage}--\pageref{lastpage}} \pubyear{2005}

\maketitle

\label{firstpage}

\begin{abstract}

Though about 80 pulsar binaries have been detected in globular
clusters so far, no pulsar has been found in a triple system in which
all three objects are of comparable mass.  Here we present predictions
for the abundance of such triple systems, and for the most likely
characteristics of these systems. Our predictions are based on an
extensive set of more than 500 direct simulations of star clusters
with primordial binaries, and a number of additional runs containing
primordial triples.  Our simulations employ a number $N_{tot}$ of
equal mass stars from $N_{tot}=512$ to $N_{tot}=19661$ and a
primordial binary fraction from $0-50\%$. In addition, we validate our
results against simulations with $N=19661$ that include a mass
spectrum with a turn-off mass at 0.8 $M_{\sun}$, appropriate to
describe the old stellar populations of galactic globular
clusters. Based on our simulations, we expect that typical triple
abundances in the core of a dense cluster are two orders of magnitude
lower than the binary abundances, which in itself already suggests
that we don't have to wait too long for the first comparable-mass
triple with a pulsar to be detected.

\end{abstract}


\section{Introduction}

Numerical simulations of star clusters routinely produce dynamically
formed triple stars, especially in the presence of primordial binaries
\citep{mcm90,heg92,mcm94}. Given that most globular clusters have a
significant population of primordial binaries, we would expect some
pulsar binaries in globular clusters to be actually part of a triple
system. In such a case, most likely all stars have comparable masses,
given the fact that encounters between binaries, and also between
binaries and single stars, are likely to expel the lightest star, and
hence subsequent encounters tend to concentrate the most massive stars
in binaries and triples. Such a system we refer to as a {\sl pulsar
triple}, i.e. a triple system in which all components have comparable
mass and one is a pulsar. (This phrase is preferable to the
alternative {\sl triple pulsar}, which would strongly suggest a triple
system in which all three components are pulsars.). The pulsar that is
detected through its radio emission can then either be one of the
inner two stars, or it can be the third star in orbit around the other
two.  The possible set of configurations is quite large: each of the
other two stars can be a neutron star, a massive white dwarf, possibly
a main sequence star, and perhaps even a black hole.  So far, no
pulsar triple with comparable masses has been discovered, among the
more than 120 pulsars found in globular clusters.

Ideally one would like to explore the formation of pulsar triples
using detailed N-body simulations which include direct integration of
the orbits and stellar evolution (therefore using one particle per
star). Unfortunately this approach is not computationally feasible
today nor it will be in the near future. In fact, even if we consider
a pulsar rich globular cluster such as Terzan 5, the population
synthesis models of \citet{iva07}, which include a simplified
treatment of the dynamics of the system, predict a total of $\approx
120$ neutron-star binaries (that is binaries with at least one neutron
star) and thus about one neutron-star triple, considering that stable
triples are about 1\% of stable binaries (e.g. see \citealt{aar01};
see also Sec.~\ref{sec:tripAB}). This immediately highlights that, in
order to gather enough statistics, one would need to run several tens
of simulations of star clusters such as Terzan 5, which has a present
mass of $3.7 \cdot 10^5 M_{\sun}$ and a number of stars $N>4 \cdot
10^5$. This is outside the capabilities of current hardware, including
the GRAPE6: the largest direct N-body simulations carried out to date
for at least a relaxation time are those by \cite{bau03} and have
N=131072 \citep[see ][]{htSCO}. Thus one is forced to introduce some
simplications to address the formation of triples stars with a
pulsar. As a first step we use equal-masses calculations as discussed
in Sec.~\ref{sec:sim}. These provide a reasonable approximation for
the dynamics of old globular clusters, where the turn-off mass is
below $1M_{\sun}$ and the typical mass ratio between two stars in the
cluster is around 2:1. Such an approach appears to be preferable to the
use of simulations that start with very massive stars and up to
several thousands particles, as is instead appropriate for the study
of young open clusters (e.g. see \citealt{fuente96,hur05}). In
addition, we validate our equal-masses results against simulations
with a mass spectrum that has a turn-off mass at $0.8 M_{\sun}$,
consistent with the age of stellar populations in galactic globular
clusters.

The aim of this paper is two-fold. First, we characterize the typical
fraction of triples that are formed dynamically in a star cluster with
a sizeable population of primordial binaries, comparing their number
density with that measured at late times in simulations that start
with primordial triples. Second, we take advantage of the fact that
triple systems with one or more pulsars have most likely a dynamical
and not primordial origin. We therefore apply our results to make an
estimate of the likely time we have to wait until the first pulsar
triple will be found as well as to predict its most likely
characteristics.

The paper is organized as follows. In the next two sections we present
an overview of observations and simulations, respectively.  In Sec.~4
we present the initial conditions for the simulations presented in
this paper.  Sec.~5 briefly summarizes the global evolution of a star
cluster with primordial binaries, extensively discussed in
\citet{heg06,tre06a,tre06b}.  In Sec.~6 we develop a simple model to
describe the formation and destruction of triples. We use the model to
interpret the observed triples fraction in our runs, attempting to
characterize its $N$-dependence. In Sec.~7 we discuss in detail the
properties of these dynamically formed triples. Sec.~8 describes the
results of some additional runs where we have started with the
presence of primordial triples. Sec.~9 presents the prospects for
detection of a triple system with a pulsar and Sec~10 sums up.

\section{Observations}

There are currently $\sim$130 pulsars known in globular clusters, the
majority of which are in binary systems \citep[for a recent review,
see][]{cr05}\footnote{See P.~Freire's cluster pulsar catalog at {\tt
    http://www2.naic.edu/$\sim$pfreire/GCpsr.html}}.  Since almost all
of these systems are millisecond pulsars (MSPs) with extraordinary
timing precisions (pulse arrival times are typically measured to
10$-$100\,$\mu$s), if they were part of a triple system, the other two
components would be easily detected, even if one of the components was
low-mass or in a long-period orbit \citep{tacl99}.

Since binary MSPs typically have orbital periods of several hours to
several days, a pulsar triple system with the MSP in the internal
orbit would likely be initially identified as a normal binary pulsar
system.  Timing observations would expose the third body in the
external orbit after a time equal to a few percent of the external
orbital period, which is likely to be less than a year.

Alternatively, a pulsar triple system could have the MSP in a
relatively long-period (months to years) external orbit around a
compact inner binary likely comprised of neutron stars, massive white
dwarfs,or a combination of both, and possibly main-sequence stars.
The MSP in such a system would be initially identified as a
long-period binary or possibly even an isolated pulsar.  Timing
observations would relatively quickly reveal the MSP's external orbit,
but unless strong orbital perturbations due to the inner binary are
present, the internal orbit might never be detected and the MSP
``companion'' will be assumed to be a single massive star or compact
object instead of a binary.

It is possible that stellar interactions could produce a triple system
where two of the members are radio MSPs.  The MSP in the external
orbit would be relatively easy to detect due to the likely small
orbital accelerations present over the short time intervals
(i.e.~hours) used for pulsar searches.  For an MSP in the compact
inner orbit, though, strong orbital accelerations would make the
initial detection of the MSP very difficult, likely requiring
specialized algorithms and extremely large amounts of computing
\citep{rce03}.

Finally one could ask what is the a-priori chance to find a triple
system with two of the three bodies being pulsar. We expect that this
probability is significantly smaller than that of finding a triple
system with just one pulsar. In fact among the $\sim$100 binaries containing
a pulsar, none is a double pulsar. This suggests that the probability
for a triple with two pulsars is at least two orders of magnitude
smaller than the probability of finding a single pulsar triple. The
expected number of triples with two pulsars ($N_{T\_pp}$) can be
estimated as:
\begin{equation}
N_{T\_pp} = \eta \frac{N_{p}}{ N_{s}} \frac{N_{B\_ps}}{ N_{B}} N_T,
\end{equation}
where $N_p$ is the number of pulsars, $N_s$ is the number of single
stars, $N_B$ is the number of binaries, $N_{B\_ps}$ is the number of
binaries with one pulsar, $N_T$ is the total number of triples of the
system and $\eta$ is the fraction of pulsars not bound to any
companion in globular clusters ($\eta \approx 0.1$ from
\citealt{freire05}). If we assume for a globular cluster $N_p \approx
10$, $N_s \approx 10^5$, $N_{B\_ps} \approx 10$, $N_B \approx 10^4$,
$N_T \approx 10^2$ we obtain $N_{T\_pp} \approx 10^{-6}$. Therefore it
is unlikely that any double-pulsar triple will be detected in globular
clusters.

\section{Simulations}\label{sec:sim}

The dense cores of globular clusters represent natural laboratories
for studying exotic stellar populations, frequently living in multiple
systems, such as blue stragglers, low mass X-ray binaries, cataclysmic
variables, millisecond pulsars, stellar and, possibly, intermediate
mass black holes. Most of these objects form through the combined
effects of stellar evolution and stellar dynamics, especially due to
interaction and evolution of binary stars, that may constitute up to
$50\%$ of the core mass of a globular cluster \citep[e.g.,
see][]{alb01,bel02,pul03}. Observational evidence for triple stars is
also starting to accumulate \citep[e.g., see][]{rab98,ste02}.

From the theoretical point of view, a detailed modeling of dense
stellar systems is extremely challenging due to the huge dynamical
range involved. Even neglecting stellar evolution and hydrodynamics
and focusing on gravitational interactions only, the local orbital
timescale of a binary star is several orders of magnitude smaller
than the global relaxation timescale (hard binaries have an orbital
period of a few hours, while the half mass relaxation time may be up
to a few billion years).

Numerical simulations with primordial binaries have thus been
performed either using approximate algorithms such as Fokker Planck or
Monte Carlo methods \citep{gao91,gie00,fre03}, that have to rely on
physical inputs on the interaction cross sections (but see
\citealt{fre05} and \citealt{gie03} for Fokker Planck codes that
treats binary interactions via direct N-Body integration), or using
direct N-Body codes but employing, until recently, only a limited
number of particles $N \approx
10^3$\citep{mcm90,heg92,mcm94,fuente96}. Studies of hierarchical
systems in open clusters have been carried out by \citet{fuente97}. A
few direct numerical simulations with $N \approx 10^4$ aimed at
studying the formation and evolution evolution of hierarchical systems
have been recently performed by \citet{aar00,aar01} and by
\citet{aar00a,aar04}.

To enhance the statistical base and remove some of the limitations of
previous direct N-body investigations, we have started a project
(\citealt{heg06,tre06a}, hereafter HTH,THH respectively; see also
\citealt{tre06b}) to study the evolution of star clusters with
primordial binaries by means of several hundred direct numerical
simulations employing up to $N_{tot}=19661$ particles. Our aim is to
provide a milestone for comparison and validations of approximate
methods and to investigate in detail the complex dynamical
interactions of multiple stellar systems, that cannot so easily be
characterized in terms of approximate treatments like Fokker Planck or
Monte Carlo methods.

In this paper we take advantage of the extensive set of numerical
simulations (more than 500) that we have performed (see HTH,THH) and
we estimate the frequency and the orbital properties of dynamically
formed triple stars. We discuss their dependence on the dynamical
state of the star cluster (pre and post-collapse), on the initial
binary ratio and on the number of particles $N$. We complement these
simulations (i) with previously unpublished runs that include a mass
spectrum appropriate to represent the collisional evolution of old
stellar populations in globular clusters (that is with a turn-off mass
below $1 M_{\sun}$) and (ii) with a few equal mass runs with a
significant population of primordial triples. Although most of our
simulations are limited to equal mass stars without stellar evolution,
we briefly discuss in the concluding section the applicability of our
results to the prediction of the properties of globular cluster triple
systems of comparable mass that include one, or more, millisecond
pulsars.

\section{Initial Conditions}

The general properties of the numerical simulations considered in this
paper to study the properties of triple systems have been presented in
detail in HTH and THH. To summarize, we use Aarseth's NBODY-6 code
\citep{aar03} considering stars of equal mass and no stellar
evolution. The initial distribution is either a Plummer model (HTH; in
this case the system is considered isolated) or King models with
concentration parameter $W_0=3,7,11$ (THH; here the effects of a
self-consistent galactic tidal field are also taken into account).
The number of objects $N$ used ranges from $256$ to $16384$; here
$N=N_s+N_b$, where $N_b$ and $N_s$ are the initial number of binaries
and single stars, respectively.  The total number of stars is
$N_{tot}=N_s+2N_b$. In our standard runs (see HTH and THH), the
primordial binary population has a fractional abundance ($f =
N_b/(N_s+N_b)$) from $0$ to $50 \%$, with an interal binding energy
distribution flat in log scale in the range $\approx 5~kT$--$680~kT$,
where $(3/2)kT$ is the mean kinetic energy per particle of the system
(the binaries being replaced by their barycenter). In addition we have
performed a few simulations with $N=8192$ and $f=20\%$ that start from
a King $W_0=7$ model (with a self consistent tidal field) and adopt an
extended binding energy range (from $5~kT$ to $10^4~kT$) for
primordial binaries. The aim of these runs is to specifically
investigate the formation of triples originated from neutron
star binaries with a degenerate companion.

In addition to these equal mass runs we also consider two tidally
limited runs ($W_0=7$ and $W_0=11$ King models) with a mass
spectrum. For these simulations, which have $N=16384$ and $f=10\%$,
the individual particle mass is drawn from an initial mass function
appropriate to study the late evolutionary stages of star clusters,
when stars are $\approx 10$~Gyr old. To build the initial mass
function for the run we first consider a \citet{ms} mass function in
the interval $[0.2,10] M_{\sun}$, and then we proceed to take into
account the effect of stellar evolution for massive stars by reducing
the mass of particles above the turn-off (assumed at $m_{toff} =
0.8~M_{\sun}$) accordingly to the prescription of \citet{hur00}.

All our results are presented using standard units \citep{hm86}
in which
$$
G = M = - 4 E_{\rm T} \equiv 1
$$ where $G$ is the gravitational constant, $M$ the total mass, and
$E_{\rm T}$ the total energy of the system of bound objects.  In other
words, $E_{\rm T}$ does not include the internal binding energy of the
binaries, only the kinetic energy of their center-of-mass motion and
the potential energy contribution where each binary is considered to
be a point mass.  The corresponding unit of time is $t_d =
GM^{5/2}/(-4E_{\rm T})^{3/2} \equiv 1$ \citep{hm86}.  For the
relaxation time, we use the following expression \citep{spitzer87}
$$
t_{rh} = \frac{0.138 N r_h^{3/2}}{\ln{(0.11 N)}}.
$$ We recall that as our simulations consider gravitational
interactions only, our results, expressed in terms of the
dimensionless units described above, can be applied to any physical
choice for the total mass and scale radius of the system. The
dependence of the triples' properties on the number of particles
employed is discussed in Sec.~\ref{sec:tripAB} and provides a way to
extrapolate the results of our simulations to a number of particles
realistic for standard globular clusters.

For reference in physical units, a globular cluster (described by a
Plummer model) with $N=3 \cdot 10^5$ stars, a total mass of $M=3 \cdot
10^5~M_{\sun}$ and a half-mass radius of $4~pc$ has a half-mass
relaxation time $t_{rh} \approx 8.5 \cdot 10^8~yr$; in this cluster a
binary, formed by equal mass stars each of mass $1~M_{\sun}$, with
binding energy of $1~kT$ has a semi-major axis of $\approx 10~AU$ and
an orbital period $p_0 \sim 20$~yrs.

\section{Global evolution}

The evolution of the system in our runs is driven by the balance of
two competing phenomena: the tendency of the core to contract
(undergoing a gravothermal collapse) and the generation of energy due
to binary-binary and binary-single interactions, that eventually halts
the core contraction and fuels the half mass radius expansion.  We can
broadly identify two phases, extensively described in HTH and THH: an
initial core radius adjustment and a quasi-steady binary burning
phase.

The first ``adjustment'' transient is characterized by the evolution
of the core toward a quasi-equilibrium configuration that will be
maintained during the subsequent binary burning phase. In this phase
we may have, depending on the initial conditions, a contraction (e.g.,
starting from a Plummer model, or from King models with $W_0 \lesssim
7$) or an \emph{expansion}, if the core is initially too small
(\citealt{fre03} and the discussion of the $W_0=11$ runs in THH). For
equal mass stars this transient lasts up to $\approx 10~t_{rh}(0)$ and
the expected core radius value at the end of this initial adjustment
is well modeled in terms of general theoretical considerations
\citep{ves94}. Essentially the efficiency of the production of energy
due to binary burning depends on the core density: if the core density
is too low, the energy production is inefficient and the core shrinks;
on the other hand a very high core density leads to an excessive
energy generation, with a consequent core expansion.

After the ``adjustment'' transient, the details of the initial
conditions are largely erased from the system and a self similar
evolution sets in: the cluster expands by keeping the ratio of the
core to half mass radius almost constant. The fuel for the expansion
is provided by the hardening and destruction of the primordial binary
population, that can sustain this phase for about $100~t_{rh}(0)$ in
isolated models (thus for a time much longer than the age of
the universe for a typical globular cluster) if the initial binary
fraction is above $10\%$.

As a result of four body interactions, basically happening in the core
of the system only, where the interaction probability is highest, a
(small) fraction of stable triple stars is formed. In the next Section
we discuss the properties of these multiple systems.

\section{Triple Formation and Abundances}\label{sec:tripAB}

Our runs start without primordial triples, but in a few relaxation
times stable triples (identified as stable in NBODY6 using a
semi-analytic approach based on the binary tides problem - see
\citealt{aar00} and \citealt{mar01}) are formed via dynamical
interactions in binary-binary encounters following ejection of one
star. We recall that stable triples cannot be formed due to three body
encounters (see, for example, \citealt{heg03}).

\begin{figure}
\resizebox{\hsize}{!}{\includegraphics{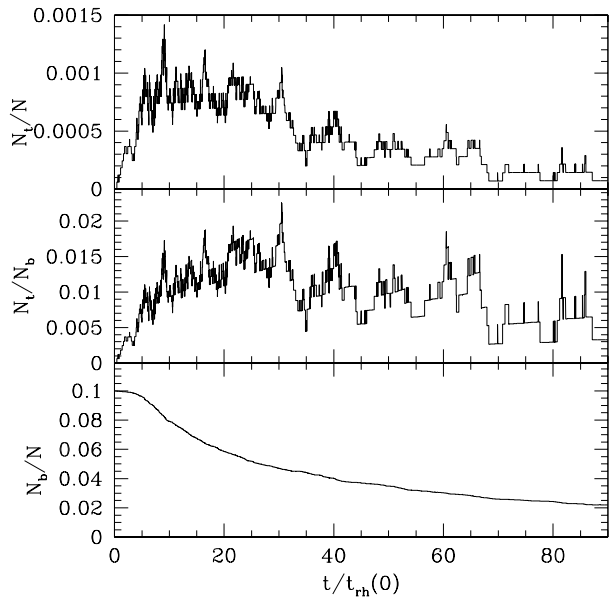}} \caption{Evolution
  of the number of triples with respect to the number of particles $N$
  (upper panel), to the number of binaries (middle panel) and number
  of binaries to the number of particles (lower panel) for a simulation
  with $N=16384$ starting from a Plummer model (isolated, no tidal
  field) and $10\%$ primordial binaries. Due to the absence of a tidal
  field the mass loss timescale is significantly longer than in the
  run presented in Fig.~\ref{fig0} and the drop in the number of
  triples at later stages of the evolution is apparent. See HTH for
  more details on the run.}\label{fig0a}
\end{figure}

\begin{figure}
\resizebox{\hsize}{!}{\includegraphics{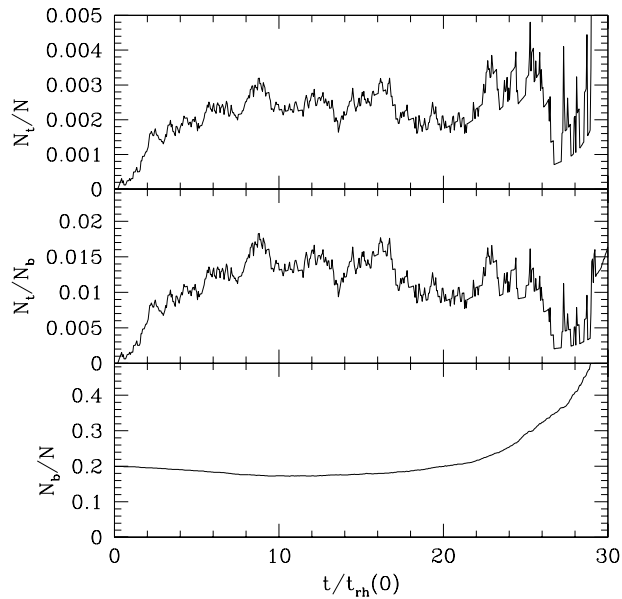}} \caption{Evolution
  of the number of triples  like in Fig. \ref{fig0a} but for a
  simulation with $N=16384$ (with a tidal field) starting from a King
  $W_0 = 7$ model and $20\%$ primordial binaries. The time is in units of
  the initial half mass relaxation time $t_{rh}(0)$. After about
  $30t_{rh}(0)$ the cluster is completely dissolved by the tidal
  field. See THH for more details on the run.}\label{fig0}
\end{figure}

After a first rapid rise of the number of triples that happens in the
first few relaxation times, a quasi stationary regime sets in and the
number of triples is approximately proportional to the number of
binaries.  This regime lasts for about 30 relaxation times, a time
longer than the age of the universe for a typical globular
cluster. Eventually, in isolated runs, as the evolution proceeds and
the reservoir of primordial binaries is being depleted, the number of
triples also drops. Fig~\ref{fig0a} illustrates the evolution of the
number of triples in an isolated run (N=16384, 10\% primordial
binaries, starting from a Plummer model) up to $\approx 90~
t_{rh}(0)$. The drop in $N_t/N$ for $t>30 t_{rh}(0)$ is clear, while
the fraction of triples to binaries declines more
slowly. Fig.~\ref{fig0} shows instead the number of triples for a run
that includes a galactic tidal field, starting from a King $W_0=7$
model with 20\% primordial binaries and N=16384.  The tidal field
destroys the cluster in about $30~t_{rh}(0)$, thus even in the last
stages of the evolution the binary fraction remains high (as single
stars preferentially escape from the cluster) and so does the triple
fraction. The asymptotic fraction of triple to single stars for our
runs with equal-mass particles is given in Tab.~\ref{tab:equalmass}.
This has been obtained by averaging the triple to single stars ratio
after the core contraption ($t>t_{cc}$) and, in the case of tidal
field runs, stopping when the number of stars in the system is reduced
below $10\%$ of its initial value.

\begin{figure}
\resizebox{\hsize}{!}{\includegraphics{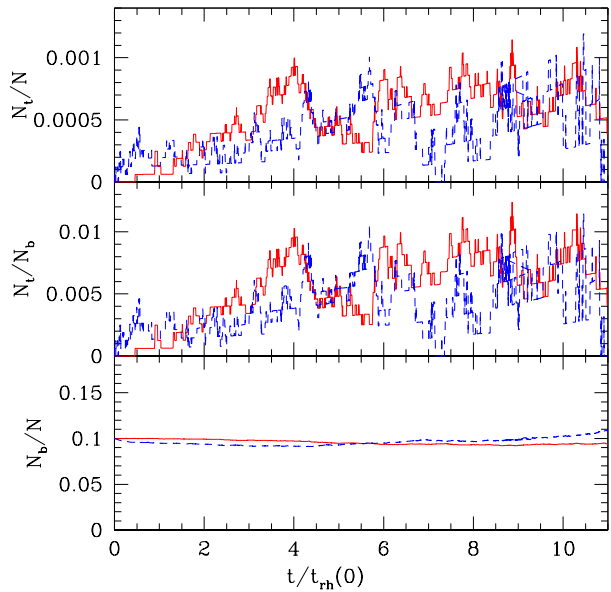}}
  \caption{Evolution of the number of triples like in Fig. \ref{fig0a}
  but for two simulations with $N=16384$ (with a tidal field) starting
  from a King $W_0 = 7$ model (red solid line) and $W_0=11$ model
  (blue dashed line) with $10\%$ primordial binaries. These
  simulations have particle masses drawn from a \citet{ms} mass
  function evolved to have a turn-off mass at $0.8
  M_{\sun}$. }\label{figMS}
\end{figure}

The evolution of the system is very similar even when a mass spectrum
is introduced in the simulations (see Fig.~\ref{figMS} and
Tab.~\ref{tab:massspect}). Within the statistical uncertainties
associated to the small number of triples present at any time in the
system ($N_t \lesssim 10$), the behavior of these runs is completely
consistent with that observed in their equal-mass particles
counterparts. These experiments with a mass spectrum confirm that
equal mass runs are a very reasonable approximationas a first step for
characterizing the properties of pulsar triples in systems where the
turn-off mass is below $1 M_{\sun}$.


The evolution of the number of triples can be understood in terms of
the balance between the formation ($r_f$) and the destruction ($r_d$)
rate for triples. In first approximation we can write locally:
\be \label{eq:pform} r_{f} = \alpha_f \cdot \rho_b^2, \ee
\be \label{eq:pdes}
r_{d} = \rho_t (\alpha_{dB} \cdot \rho_b +\alpha_{dS} \cdot \rho_s),
\ee
where $\rho_s,\rho_b,\rho_t$ are the density of singles, binaries and
triples respectively, while the $\alpha_i$ are numerical coefficients
that depends on the cross section for the process considered. 

\begin{figure}
\resizebox{\hsize}{!}{\includegraphics{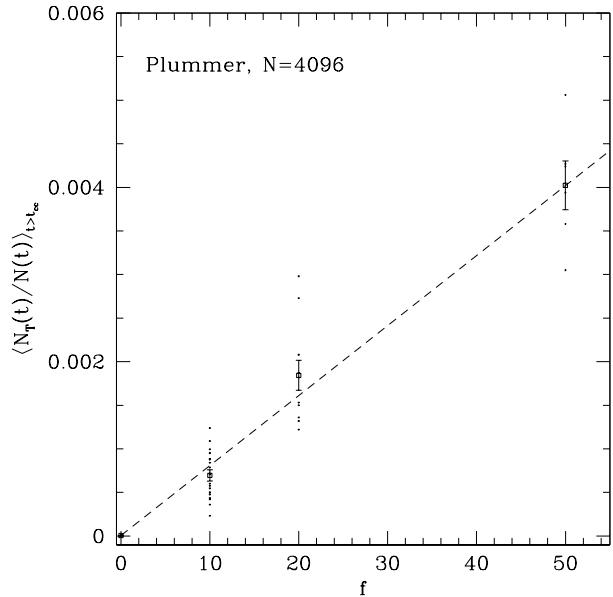}} \caption{Fractional
  abundance of triples averaged after the core contraction for
  isolated runs starting from a Plummer model and different primordial
  binary ratios. Each dot represents a different run, while the
  squares, with a $1\sigma$ error bar, are the average values of the
  fractional abundances of triples associated with a given primordial
  binary ratio $f$. The number of triples appears to be linearly
  proportional to the primordial binary fraction.}\label{fig5}
\end{figure}

With the aid of Eqs.~(\ref{eq:pform}-\ref{eq:pdes}) we can explain the
behavior of Fig.~\ref{fig0}. During the initial core contraction phase
the formation rate $r_f$ is increasing as binaries accumulate in the
core, while the destruction rate $r_d$ is smaller due to the relative
absence of triples, whose numbers therefore increase. Around core
collapse the formation rate reaches its peak and in the mean time an
equilibrium regime sets in where the density of triples can be
obtained by equating the formation and destruction rate ($r_f=r_d$):
\be \label{eq:rhoT} \rho_t = \frac{\alpha_f \cdot
\rho_b^2}{\alpha_{dB} \cdot \rho_b +\alpha_{dS} \cdot \rho_s}.  \ee
We can further assume, based on geometrical arguments, that the cross
section for a binary-triple encounter is greater than that of a
triple-single encounter. In addition we recall that after core
collapse in simulations starting with $f \lesssim 10\%$ the central
binary density at least equals that of single stars (e.g. see
Fig.~21 in HTH).  This leads to a simplification of
Eq.~(\ref{eq:rhoT}):
\be \label{eq:rhoTsempl}
\rho_t = \frac{\alpha_f \cdot \rho_b^2}{\alpha_{dB} \cdot \rho_b}= \frac{\alpha_f}{\alpha_{dB}} \rho_b.
\ee
This theoretical prediction is nicely confirmed in Fig.~\ref{fig5},
where the abundance of triples after core collapse for runs starting
from $N=4096$ is shown as a function of the fraction of primordial
binaries: here the structural properties of the system are nearly
identical for $f \gtrsim 10\%$ (see Figs. 17 \& 18 in HTH) and there
is a linear dependence of $N_t$ on $f$.

Finally at later stages of the evolution of the system (i.e. well
after the end of core-contraction phase), $\rho_b$ in the core drops
as binaries are being depleted, and the number of triples also
decreases.

\begin{figure}
\resizebox{\hsize}{!}{\includegraphics{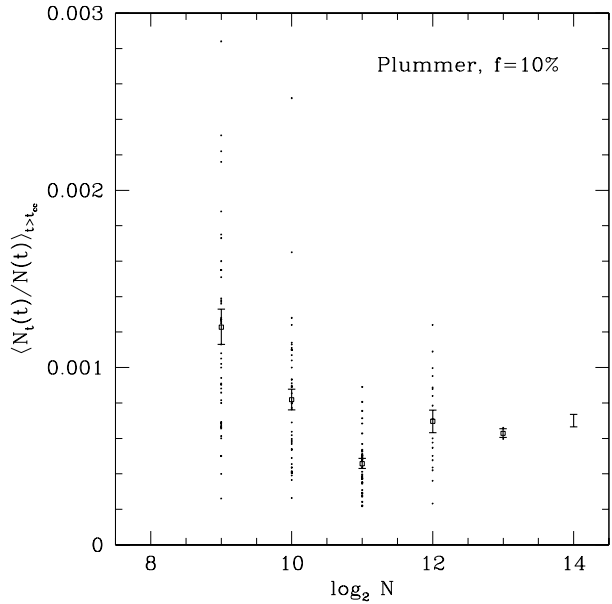}} \caption{Fractional
  abundance of triples averaged after the core contraction for
  isolated runs starting from a Plummer model with $10\%$ primordial
  binaries and different numbers of particles. Symbols are as in
  Fig.~\ref{fig5}}\label{fig1}
\end{figure}

\begin{figure}
\resizebox{\hsize}{!}{\includegraphics{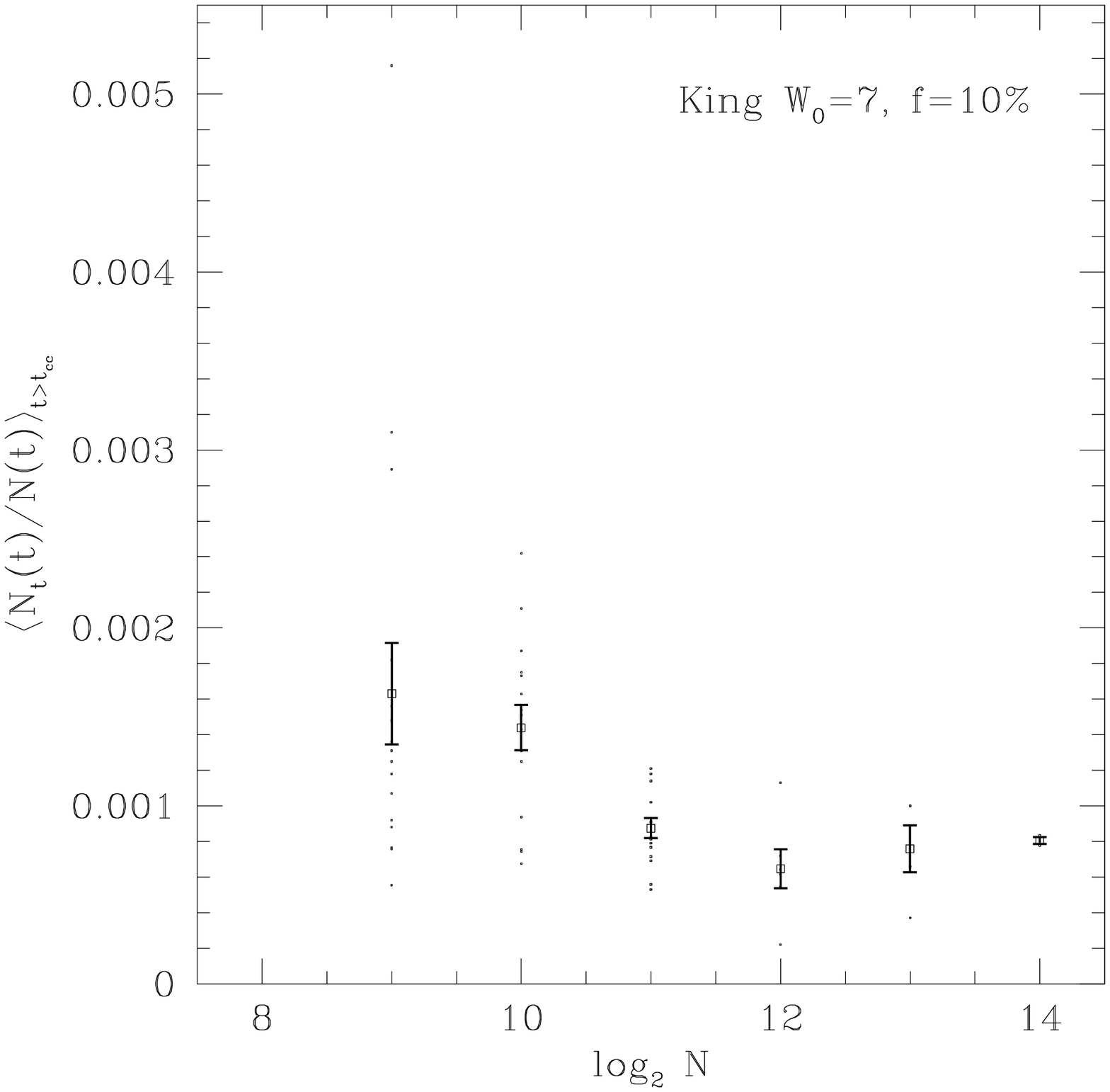}} \caption{Like
  Fig.~\ref{fig1} but for simulations with a tidal field starting from
  a King $W_0=7$ model.}\label{fig2}
\end{figure}

\begin{figure}
\resizebox{\hsize}{!}{\includegraphics{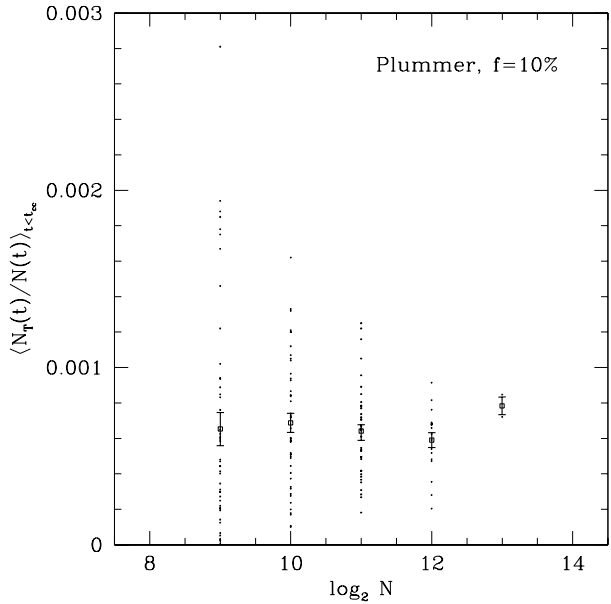}} \caption{Fractional
  abundance of triples averaged before the core contraction for
  isolated runs starting from a Plummer model with $10\%$ primordial
  binary and different number of particles.}\label{fig3}
\end{figure}

\begin{figure}
\resizebox{\hsize}{!}{\includegraphics{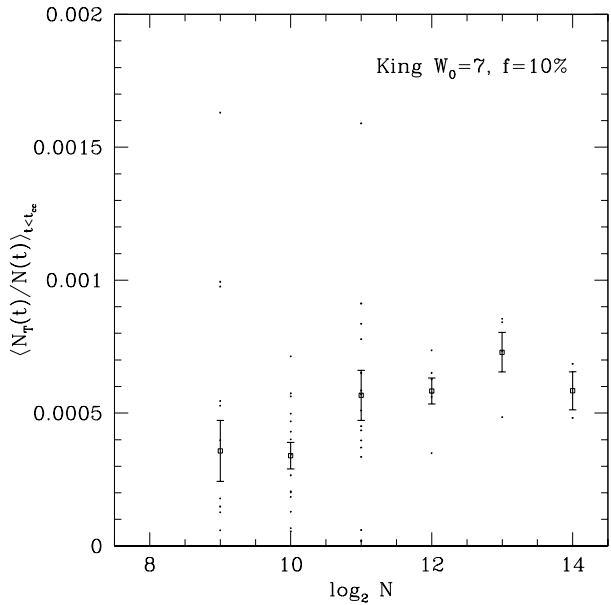}} \caption{Like
  Fig.~\ref{fig3} but for simulations with a tidal field starting from
  a King $W_0=7$ model.}\label{fig4}
\end{figure}

The fractional abundance of triples as a function of the number of
particles used, for runs starting with $f=10\%$, either from a Plummer
or from a King $W_0=7$ model, is depicted in
Figs.~\ref{fig1}-\ref{fig4}. Especially for low N runs, before the
core collapse the triple abundance is lower than after it. This is
because the time for core collapse (in units of the relaxation time)
increases with N (see HTH, Fig. 17), so that when N is low enough the
system has not yet reached a balance between the formation and
destruction rate of triples. In the runs starting from a King model we
observe a marginally higher ratio of triples. This is due to the fact
that the runs starting from a King model have a tidal field (see THH),
so that the system steadily loses stars in its outskirts and the ratio
of triples over total number of stars is enhanced in the post-collapse
phase with respect to isolated runs starting from a Plummer model.


The $N$ dependence of the number of triples at fixed initial binary
ratio is more difficult to characterize than the dependence on the
binary ratio at fixed $N$. In fact, while in the latter case the
structural properties of the cluster are fixed, by varying $N$ the
core to half mass ratio is changing \citep[see ][ and HTH,
THH]{ves94}, so that the analysis in terms of
Eqs.~(\ref{eq:pform}-\ref{eq:pdes}) is complicated by the necessity to
integrate over the density profile and by the N-dependence of the
$\alpha$ coefficients that describe the efficiency of the different
formation/destruction channels. Empirically we can also note the
intrinsic large scatter in the measured triples abundances (see the
dots representing individual runs in Figs.~\ref{fig1}-\ref{fig4}), so
that it is hard to draw a firm conclusion. Tentatively, the $N$
dependence is logarithmic at most, especially at lower N (in fact the
core radius decreases approximately as $1/log(0.1 \cdot N)$). However
for the last three points (N=4096,8192,16384) the data are consistent
with a constant value. In fact the decrease in the core radius is
compensated by a higher central density and by an increased number of
binaries over singles in the core (see Fig.~21 in HTH), so that the
fractional abundance of triples should stay approximately constant.

The evolution proceeds in a qualitatively similar way in our runs
initialized with an extended binary binding energy range (up to
$10^4~kT$ ). The only difference is that, all other conditions being
equal, triples formation is enhanced by a factor between 1.5 and 2
when harder primordial binaries are present. In fact, binaries that
have binding energies above $\approx 700~kT$ are dynamically inert and
behave essentially like singles during gravitational interactions with
other stars. Therefore the encounter of one of these ultra-hard
binaries with a regular (hard) binary can lead to an exchange
encounter where one component of the regular binary is replaced by the
ultra-hard pair, much like during a single-regular binary
interaction. A more efficient production channel for stable triples is
therefore available in these simulations. Therefore neutron star
binaries with a degenerate companion and with short orbital periods
have an enhanced probability of ending up being in a triple system
compared to main sequence binaries.

To summarize, in a typical globular cluster ($N=3 \cdot 10^5$) we
expect an order of at least 100 triples if we conservatively assume a
binary fraction of 10\% (with a standard binding energy range, up to
contact main sequence binaries) and $N_t/N_b \approx 5 \cdot 10^{-3}$
(this accounts for the effect of a mass spectrum and tidal field in
setting $N_T/N_b$; see Fig.~\ref{figMS}). The number approximately
doubles by adopting an extended binding energy range that includes
binding energies reached by neutron star binaries with a degenerate
companion.

\begin{table}
\caption{Asymptotic triple fraction in our equal mass runs. The first
column reports the number of particles $N$, the second the initial
density profile (Isolated Plummer model - $Pl$ or tidally limited king
model - $W_0$), the binary fraction $f$ is in the third column. The
fourth and fifth entries are the average asymptotic triple fraction
and its standard deviation, computed over $N_{runs}$ realizations
(last entry).}\label{tab:equalmass}
\begin{tabular}{rccccc}
\hline
N & Model & f & $\langle N_t/N \rangle$ & $\sigma_{N_t/N}$ & $N_{runs}$ \\ 
\hline
  512  &    Pl  &   0 & 0.80E-04 & 0.11E-03 &    48 \\
  512  &    Pl  &  10 & 0.12E-02 & 0.68E-03 &    48 \\
  512  &    Pl  &  20 & 0.22E-02 & 0.81E-03 &    47 \\
  512  &    Pl  &  50 & 0.45E-02 & 0.14E-02 &    25 \\
 1024  &    Pl  &   0 & 0.35E-04 & 0.31E-04 &    49 \\
 1024  &    Pl  &  10 & 0.82E-03 & 0.39E-03 &    46 \\
 1024  &    Pl  &  20 & 0.19E-02 & 0.62E-03 &    49 \\
 1024  &    Pl  &  50 & 0.47E-02 & 0.13E-02 &    47 \\
 2048  &    Pl  &   0 & 0.13E-04 & 0.16E-04 &    46 \\
 2048  &    Pl  &  10 & 0.46E-03 & 0.16E-03 &    35 \\
 2048  &    Pl  &  20 & 0.17E-02 & 0.44E-03 &    39 \\
 2048  &    Pl  &  50 & 0.41E-02 & 0.93E-03 &    37 \\
 4096  &    Pl  &   0 & 0.50E-05 & 0.37E-05 &    7  \\	
 4096  &    Pl  &  10 & 0.70E-03 & 0.27E-03 &    18 \\	
 4096  &    Pl  &  20 & 0.18E-02 & 0.54E-03 &    11 \\	
 4096  &    Pl  &  50 & 0.40E-02 & 0.62E-03 &    6 \\	
 8192  &    Pl  &   0 & 0.18E-05 & N/A      &    1 \\	
 8192  &    Pl  &  10 & 0.63E-03 & 0.31E-04 &    2 \\	
 8192  &    Pl  &  20 & 0.17E-02 & N/A      &    1 \\	
 16384 &    Pl  &  10 & 0.71E-03 & N/A      &    1 \\	
\hline  
  512 &    $W_0=7$  &  10 & 0.16E-02 & 0.11E-02 & 16 \\
 1024 &    $W_0=7$  &  10 & 0.14E-02 & 0.50E-03 & 15 \\
 2048 &    $W_0=7$  &  10 & 0.87E-03 & 0.22E-03 & 15 \\
 4096 &    $W_0=7$  &  10 & 0.65E-03 & 0.27E-03 & 6  \\
 8192 &    $W_0=7$  &  10 & 0.76E-03 & 0.26E-03 & 4  \\
16384 &    $W_0=7$  &  10 & 0.81E-03 & 0.28E-04 & 2  \\
16384 &    $W_0=7$  &  20 & 0.15E-02 & N/A & 1  \\
16384 &    $W_0=11$ &  10 & 0.54E-03 & N/A & 1  \\
16384 &    $W_0=11$ &  20 & 0.22E-02 & N/A & 1  \\

\hline
\hline
\end{tabular}
\end{table}

\begin{table}
\caption{Asymptotic triple abundance for tidally limited runs with a
Miller \& Scalo mass spectrum. Columns as in
Tab.~\ref{tab:equalmass}.}\label{tab:massspect}
\begin{tabular}{rccccc}
\hline
N & Model & f & $\langle N_t/N \rangle$ & $\sigma_{N_t/N}$ & $N_{runs}$ \\ 
\hline
  16384  &    $W_0=3$ &  10 & 0.98E-03 & N/A &    1 \\
  16384  &    $W_0=7$ &  10 & 0.68E-03 & N/A &    1 \\
  16384  &    $W_0=11$ & 10 & 0.61E-03 & N/A &    1 \\
\hline
\hline
\end{tabular}
\end{table}

%
%
%

\section{Properties of Newly Formed Triples}

While for the majority of the standard simulations in our sample we
recorded only the number of stable triples, we have a small number of
high resolution (N=16384) simulations with complete information on the
orbital properties of triples, saved every $10~t_d$. In addition,
all the extended binding energy range simulations have recorded
information of the orbital properties of multiple systems.

\begin{figure}
\resizebox{\hsize}{!}{\includegraphics{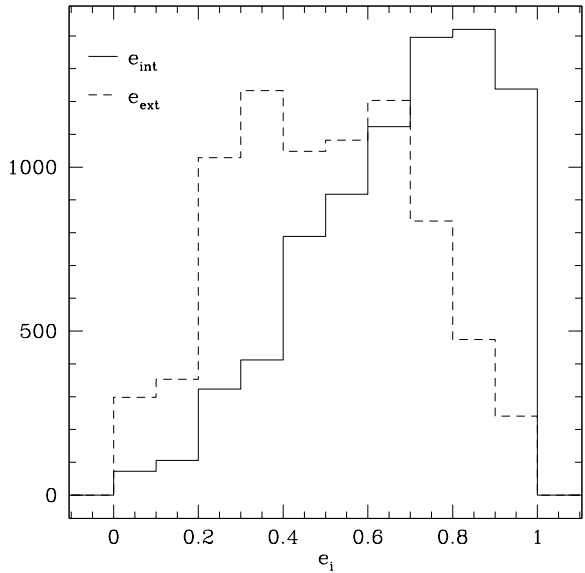}} \caption{Internal
  ($e_{int}$) and external ($e_{ext}$) eccentricity distribution for
  triples measured in a series of simulations with a tidal field
  starting from King $W_0=3,7,11$ models with $20\%$ primordial
  binaries.}\label{fig:ecc}
\end{figure}

For a sample of 3 simulations with N=16384 and $f=20\%$,
Fig.~\ref{fig:ecc} shows the distribution of internal and external
eccentricities. The inner eccentricity has a distribution peaked at
high eccentricities which derives (i) from the input thermal
distribution of the eccentricity for primordial binaries and (ii) from
\citet{kozai} perturbations induced by the outer body, which lead to
growth of the inner binary eccentricity \citep{aar01,aar00}. Note that
our simulations are limited to Newtonian dynamics, so we are missing
relativistic corrections, which may alter the orbits of the inner
binary \citep{aar07}. The outer component of a stable triple is
instead biased toward more circular orbits. This means that there is a
selective disruption of triple systems with eccentric outer orbits.

\begin{figure}
\resizebox{\hsize}{!}{\includegraphics{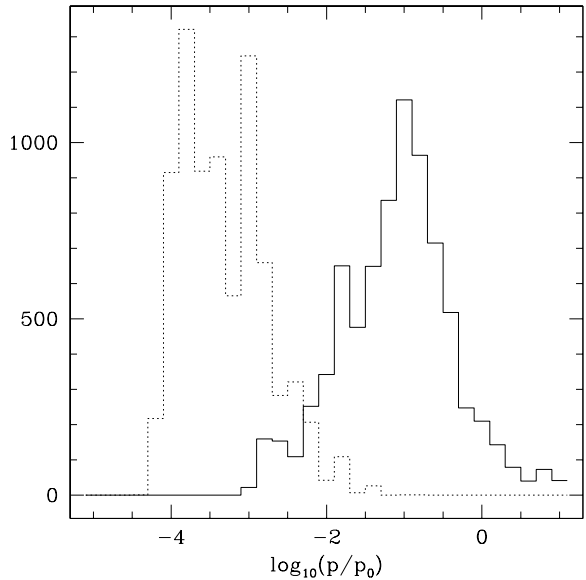}} \caption{Internal
  (dotted) and external (solid) period distribution - in units of the
  period for a 1kT binary ($p_0$) - for triples measured in a series
  of simulations with a tidal field starting from King $W_0=3,7,11$
  models with $20\%$ primordial binaries and $N=16384$.}\label{fig:period}
\end{figure}

For the same simulations the period distribution is reproduced in
Fig.~\ref{fig:period}. Here we can note that the internal component
has an orbital period between two and three orders of magnitude
shorter than the outer component. The internal orbit corresponds on
average to a binary with binding energy of a few hundreds of $kT$
while the external orbit has a binding energy of a few $kT$. The
distribution of orbital periods has a lower limit at
$\log{(p/p_0)}=-4.25$, where $p_0$ is the orbital period of a 1kT
binary. This lower limit corresponds to a binding energy of $\approx
700kT$, i.e. a quasi-contact binary with $1 M_{\sun}$ main sequence
components.

The dynamical interactions in the star cluster lead to a fast
destruction of triples with external binding energy below $1kT$; the
destruction of the typical triples found in our simulations (with
external binding energy of $5kT$) happens on a timescale of about 10
half mass relaxation times.

\begin{figure}
\resizebox{\hsize}{!}{\includegraphics{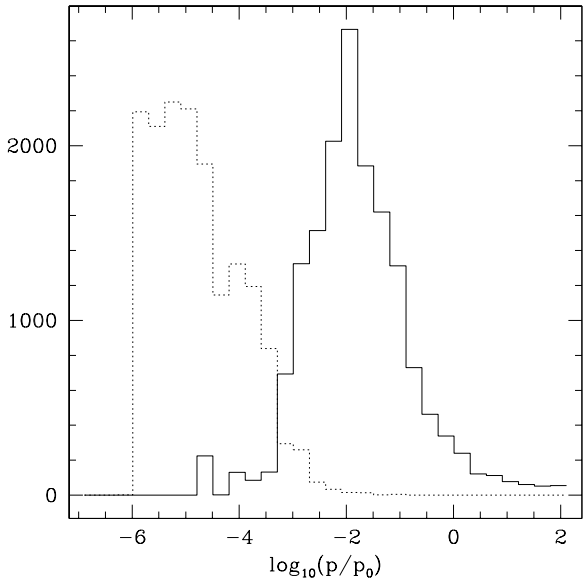}} \caption{Internal
  (dotted) and external (solid) period distribution - like in
  Fig.~\ref{fig:period} - for triples measured in a series
  of simulations with a tidal field starting from King $W_0=7$
  models with $20\%$ primordial binaries, $N=8192$ and an extended
  initial binding energy range ($[5~kT:10^4~kT]$), representative for
  neutron star binaries with a degenerate companion.}\label{fig:period_ext}
\end{figure}

The eccentricity distribution in our extended $kT$ range runs is
similar to that plotted in Fig.~\ref{fig:ecc}, showing only a
marginally less marked circularization of the outer orbit. The
distribution of the orbital periods is instead different (see
Fig.~\ref{fig:period_ext}). Stable triples are composed preferentially
of an inner ultra hard binary ($E_b \gtrsim 500~kT$) and an external
component that, on average, has a period only marginally shorter than
in the standard runs.

The limited statistics in our simulations do not allow us to
empirically constrain the N-dependence of the dynamical properties of
triples. On the basis of theoretical modeling we do not expect
significant variations with N for single mass star clusters, so that
the orbital properties discussed here should be representative for
simulations of globular clusters with a realistic number of
particles. However one important caveat must be made as there is no
guarantee that these properties are representative for the triple
population formed in a realistic model that includes a mass spectrum
and stellar evolution.

\section{Primordial Triples}

If a star cluster originates with a (small) population of primordial
triples, how does the number of surviving primordial triples after
several billions of years compare with the number of dynamically
formed triples? To answer this question we have performed a few
simulations with $N=4096$ starting from an isolated Plummer model and
with a number of primordial triples ($f_{trip}=5\%$ and
$f_{trip}=10\%$, where $f_{trip}$ is the initial number fraction of
triples). These runs are, to the best of our knowledge, the first
attempt to investigate the dynamical evolution of a star cluster with
a number of particles and of primordial triples greater than the few
hundreds particles used by
\citet{ber06}.

\begin{figure}
\resizebox{\hsize}{!}{\includegraphics{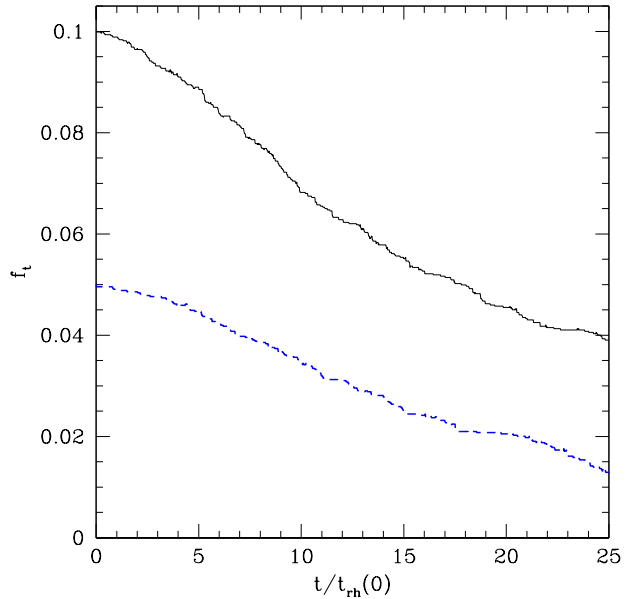}} \caption{Evolution
  of the fraction of triples $f_{trip}$ in two runs with $N=4096$
  starting from an isolated Plummer model and $f_{trip}=5\%,10\%$. The
  destruction rate is qualitatively well described by
  Eq.~\ref{eq:pdes}. }\label{fig:Ntrip}
\end{figure}

We started these runs by initializing the inner binary components with
the same procedure adopted in Heggie, Trenti \& Hut (2006), i.e. in an
energy range $[5:680]~kT$. An external component has then been added
using the triples initialization subroutine within NBODY6 (Aarseth
2003), with an energy in the range $[0.1:100]~kT$ when the external
energy is defined as $2m/a_{ext}$ with $m$ being the single particle mass and
$a_{ext}$ being the external semi-axis. This means that the softest
external component has a semi-axis 100 times larger than the softest
inner binary. The evolution of the number of triples is shown in
Fig.~\ref{fig:Ntrip}: after a slow start (due to the initial large
core, and therefore low density of triples in the core) primordial
triples are steadily burned in the first $15~t_{rh}(0)$ with a rate
approximately proportional to the number of remaining triples (note
that if multiplied by a factor 2 the curve for $f_{trip} =5\%$
reproduces closely the behavior of the curve for
$f_{trip}=10\%$). This agrees well with the expectation based on
Eq.~\ref{eq:pdes}.

At later times the destruction rate slows down slightly in
Fig.~\ref{fig:Ntrip} due to the expansion of the half mass radius,
which increases the instantaneous relaxation time with respect to the
initial one used in the figure. At about $25~t_{rh}(0)$, a time larger
than the typical age of a galactic globular cluster, the number of
primordial triples is higher than the number of dynamically formed
triples from $10-20 \%$ primordial binaries runs, if the initial
fraction of triples is higher than about one percent. If we add also
primordial binaries in runs with primordial triples, the destruction
rate of triples is slightly enhanced, as expected from
Eq.~\ref{eq:pdes} (we have verified this with a N=1024 run with
$f_{trip}=3\%$ and $7\%$ primordial binaries).

\section{Detection of a triple system with one pulsar}

As described in Sec.~4, for typical cluster parameters and $1
M_{\sun}$ stars, a 1kT binary has a semi-major axis of $\sim$10\,AU,
and an orbital period of $p_0$$\sim$20\,yrs.  Such an orbit is on the
high-end of the external orbit distribution as shown in Figure~9.  For
an MSP in such a 1kT orbit, the instantaneous orbitally induced period
derivative would be $\dot P_{orbit}$$\sim$10$^{-14}$ (compared to a
typical intrinsic $\dot P_{\rm PSR}$$\sim$10$^{-20}$, and 1-yr timing
accuracies of $\Delta\dot P$$\sim$10$^{-22}$), and would be identified
within a couple months of timing observations\footnote{This time is
longer than what would be estimated by simply considering the six
orders of magnitude difference in the accuracy of the 1-yr timing
vs. the required accuracy for detection of a triple system. This is
because there are many degeneracies (for example a positional offset)
that can mask as a period derivative and that require a few months of
observations before being excluded.}.  Shorter orbital periods would
cause significantly larger period derivatives since $\dot
P_{orbit}\propto p^{-4/3}$, where $p$ is the orbital period.  There is
no evidence for such orbitally induced accelerations in any of the
$\sim$60 globular cluster binary pulsars that have been or are
currently being timed.

Over the past several years, intensive globular cluster pulsar
searches have uncovered numerous systems that are almost certainly due
to exchange interactions.  These systems include at least 10 recycled
pulsars in highly eccentric ($e > 0.25$) orbits with periods between
$\sim$1$-$30\,days \citep[e.g.~][]{fgri04,rhs+05}.  Such systems are
very similar to the external orbits found in the triples described in
Figure~9.  However, precise timing observations rule out the existence
of internal orbits composed of at least one main-sequence star to a
high degree of confidence.  Surveys have also uncovered another class
of exchange products, the 3 pulsar$-$``main-sequence'' binaries
NGC6397~A, Terzan~5~P, and Terzan~5~ad \citep{dpm+01,rhs+05,hrs+06}.
These systems have circular orbits of duration 0.3$-$1.5\,days and
strange ``bloated'' companions that cause irregular eclipses.  The
low-eccentricity orbits were likely produced via tidal
circularization.  The erratic timing and irregular eclipses could be
caused by an internal orbit,  but the small orbital separations
(2$-$6 $R_{\sun}$) and mass-function constraints indicate likely
main-sequence dwarf companions.

The simulations we present suggest there should be approximately 100
times more binaries than triples in a typical globular cluster (see
Figure 2).  Given that we know of $\sim$100 globular cluster MSP
binaries and have already one confirmed triple system, the strange
MSP-WD-planet system in M4 \citep{tacl99}, the current data are
roughly consistent with the simulations.  The next generation of radio
telescopes, like the Square Kilometer Array, should uncover hundreds
of new globular cluster MSPs, greatly increasing our chances of
finding a pulsar triple system comprised of three stars of comparable
mass.

\section{Discussion}

In this paper we have presented a systematic investigation of the
frequency of dynamically formed triples for an extensive set of direct
simulations starting with a significant fraction of primordial
binaries. 

We have shown that on a timescale of a few relaxation times the
abundance of triples reaches an equilibrium value that depends
linearly on the primordial density of binaries and that is about two
orders of magnitude smaller. Simulations including a tidal field
present a relatively larger fraction of triples ($N_t/N$) with respect
to isolated runs with similar initial conditions. The presence of a
mass spectrum tends instead to reduce the triple fraction.

Stable triple stars in our standard runs, representative for main
sequence stars, are primarily formed due to four body encounters that
lead to the escape of a single star. The formation mechanism
influences the properties of these systems, which are primarily made
of a hard inner binary, and which in a typical globular cluster would
have an average orbital period of a few days, accompanied by an
external star with a period on average between 100 and 1000 times
longer. The inner component is also, on average, on a slightly more
eccentric orbit than the outer member of the triple. In fact eccentric
outer orbits are preferentially perturbed by gravitational encounters
or break up the inner binary.

Neutron stars binaries with a degenerate companion can reach binding
energy well above the $700~kT$ limit of two main sequence stars. To
quantify the rate of formation of stable triples involving two
degenerate stars we have carried out a series of simulations with
binding energy distributed up to $10^4~kT$. In these runs we observe
an enhanced triple abundance, as an ultra-hard binary made of two
degenerate stars essentially behaves like a single body during
gravitational encounters and can efficiently interact with a standard
binary to form a triple via an exchange-like encounter. In a typical
globular cluster, the inner orbital period may in this case be of the
order of a few hours, while the outer component of the system is
expected to be of the order of a few years, much like in the case of
triples with main sequence stars.

Triple stars with at least one pulsar component are expected to have a
dynamical origin, as it is highly unlikely that a pre-existing triple
system can avoid being disrupted by the Supernova explosion that forms
the pulsar. Therefore we expect that the properties of pulsar triples
are in agreement with those derived by runs starting with primordial
binaries only, where all the triples have been formed trough stellar
encounters (though a single pulsar could become a member of a
primordial triple via an exchange process). To date, $\sim$100 pulsar
binaries are known in the globular cluster system (Camilo \& Rasio
2005).  Based on our idealized simulations, we expect that the
discovery of a pulsar in a triple system of comparable masses is
likely to happen soon. However, we note that the majority of the known
pulsar binaries have companions of mass $\sim$10\% of the pulsar mass
rather than $\sim 1~M_{\sun}$. How this difference affects our
conclusions is currently unknown and will demand more detailed studies
with a spectrum of stellar masses beyond that used in this study.
 
Higher order hierarchical systems, such as quadruplets and quintuplets
are also occasionally observed during our runs, and their average
number density is about two orders of magnitude smaller than that of
triples. This makes a more detailed characterization of their
properties extremely challenging, at least for simulations with only a
limited number of particles like ours. A survey aimed at quantifying
the observed fraction of multiple stars in globular clusters would
allow us to understand if the observed frequency of triples and higher
order systems is consistent with dynamical production from primordial
binaries. Our investigation suggests that a triple to binary ratio up
to $\approx 2\%$ is consistent with a purely dynamical origin. If the
observed number of triples is higher, then primordial triples have to
be introduced in the theoretical modeling, as investigated for small N
open clusters ($N=182$) by \citet{ber06}. \\

\noindent {\bf{Acknowledgements}}

We thank the referee for a careful reading of the manuscript and John
Fregeau for useful suggestions. MT was supported in part by NASA award
HST-AR.10982.

\bsp

\label{lastpage}

\end{document}